\documentclass[twocolumn]{aastex7}
\usepackage{xcolor}
\usepackage{hyperref}
\usepackage{subfiles}
\usepackage{amsmath}

\begin{document}

\title{Comprehensive Analysis of Optical brightness and Color Variability of Blazars in the ZTF Survey DR22}
\correspondingauthor{Qi Yuan; Lang Cui}
\email{Emails: yuanqi@cho.ac.cn (QY)
cuilang@xao.ac.cn (LC)} 
 
\author[0000-0003-4671-1740]{Qi Yuan}\email{yuanqi@cho.ac.cn}
\affiliation{Changchun Observatory, National Astronomical Observatories, Chinese Academy of Sciences, Changchun 130117, People's Republic of China}
\author[0000-0001-8221-9601]{Xin Wang}\email{wangxin2019@xao.ac.cn}
\affiliation{Xinjiang Astronomical Observatory, Chinese Academy of Sciences, 150 Science 1-Street, Urumqi 830011, People’s Republic of China}
\affiliation{School of Astronomy and Space Science, University of Chinese Academy of Sciences, Beijing 100049, China}
\author[0000-0002-0187-6173]{Meng Zhang}\email{zhangmeng@qfnu.edu.cn}
\affiliation{School of Cyber Science and Engineering, Qufu Normal University, Qufu 273165,  People's Republic of China}
\author[0000-0002-4521-6281]{Wenwen Zuo}\email{wenwenzuo@shao.ac.cn}
\affiliation{Shanghai Astronomical Observatory, Chinese Academy of Sciences, 80 Nandan Road, Shanghai 200030, People’s Republic of China}
\author{Yan Xu}\email{xuy@cho.ac.cn}
\affiliation{Changchun Observatory, National Astronomical Observatories, Chinese Academy of Sciences, Changchun 130117, People's Republic of China}
\author{Chunguo Wu}\email{wucg@jlu.edu.cn}
\affiliation{Key Laboratory of Symbolic Computation and Knowledge Engineering of Ministry of Education, Jilin University, Changchun, 130012, People's Republic of China}
\author[0000-0002-8315-2848]{Ming Zhang}\email{zhangm@xao.ac.cn}
\affiliation{Xinjiang Astronomical Observatory, Chinese Academy of Sciences, 150 Science 1-Street, Urumqi 830011, People’s Republic of China}
\affiliation{Key Laboratory of Radio Astronomy and Technology, CAS, A20 Datun Road, Chaoyang District, Beijing 100101, People’s Republic of China}
\affiliation{Xinjiang Key Laboratory of Radio Astrophysics, 150 Science 1-Street, Urumqi 830011, People’s Republic of China}
\author[0000-0001-9815-2579]{Xiang Liu}\email{liux@xao.ac.cn}
\affiliation{Xinjiang Astronomical Observatory, Chinese Academy of Sciences, 150 Science 1-Street, Urumqi 830011, People’s Republic of China}
\author[0000-0003-0721-5509]{Lang Cui}\email{cuilang@xao.ac.cn}
\affiliation{Xinjiang Astronomical Observatory, Chinese Academy of Sciences, 150 Science 1-Street, Urumqi 830011, People’s Republic of China}
\affiliation{Key Laboratory of Radio Astronomy and Technology, CAS, A20 Datun Road, Chaoyang District, Beijing 100101, People’s Republic of China}
\affiliation{Xinjiang Key Laboratory of Radio Astrophysics, 150 Science 1-Street, Urumqi 830011, People’s Republic of China}

\begin{abstract}
This study conducts a comprehensive analysis of brightness and color variability in blazars, utilizing over six years of quasi-simultaneous g-band and r-band data from {1149} sources in the ZTF Data Release 22 (DR22), including {589} BL Lacs and {560} FSRQs.
We quantify the amplitude of variability and the fractional root mean square (rms) variability for each source and statistically assess the overall and short-term color behaviors across different subclasses; examine the distribution of brightness variability characteristics across different blazar types and investigate how the extent of variability correlates with color trends.
We found BL Lacs tend to exhibit a BWB (bluer when brighter) trend, while FSRQs display a RWB (redder when brighter) trend; 
{BL Lacs with negligible host-galaxy contamination exhibit a BWB trend fraction of 14.7\% (68/462) compared to 2.3\% (11/462) for RWB trend, while FSRQs show 8.8\% (49/560) BWB trend versus 14.1\% (79/560) RWB trend.}
{By statistically investigating how color behavior depends on brightness state across different timescales, we find that brighter states in both BL Lacs and FSRQs are more likely to exhibit BWB trend.}
Our results also show that BL Lacs with a BWB trend exhibit higher variability than those with a RWB trend, whereas FSRQs with a RWB trend display significantly greater variability than those with a BWB trend. 
{These results suggest that blazar color variability depends jointly on source type, brightness state, and variability amplitude, highlighting the complexity of color evolution in blazars.}

\end{abstract}


\section{introduction}
\label{sec:intro}

{Blazars are among the most violently variable active galactic nuclei (AGN), showing strong flux variations over timescales ranging from years to hours, and in some cases even minutes~\citep{Bottcher.19.galaxies}. These brightness variations are often accompanied by spectral or color changes, making the relation between color index and brightness an important observational diagnostic of the underlying radiation processes.}

{Extensive investigations of individual blazars and small samples have revealed a rich diversity of color behaviors. 
These source-by-source results suggest that blazar color variability is complex.}
Earlier studies of blazar optical color variability have identified a type-dependent behavior, known as the dichotomy of color behavior. For flat-spectrum radio quasars (FSRQs), the optical color behaviors to become redder as brightness increases (i.e., redder when brighter; RWB), likely due to enhanced jet radiation and diminished blue emission from the accretion disk~\citep[and references therein]{Gu.06.aa, Bonning.12.apj}.
In contrast, BL Lacs often show a bluer color as their brightness increases (i.e., bluer when brighter; BWB), possibly due to reduced cooling of high-energy electrons and the absence of significant accretion disk radiation~\citep{Massaro.98.mn, Villata.02.aa, Vagnetti.03.apj, Gu.11.aa, Gaur.12.mn}.
These trends are considered typical optical variability signatures for FSRQs and BL Lacs.
{Although a type-dependent dichotomy in optical color behavior has often been reported for blazars, this trend should be understood as a statistical tendency rather than a strict classification, since the observed color behavior can also vary with brightness state, variability amplitude, and timescale.}
{Based on detailed case studies of individual sources, such as 3C~279 \citep{Isler.17.apj}, which motivated a continuous physical framework, and S5~0716+714~\citep{Wu.2007.aj, Poon.09.apjs, Chandra.11.apj, Hu.2014.mn, Hong.17.aj}, which exhibits diverse color behaviors across different timescales, it has become increasingly clear that blazar optical color variability cannot be fully described by a simple dichotomous classification.}

{Previous studies were often based on relatively small samples, typically ranging from a few to several dozen sources, and on comparatively short monitoring time baselines, frequently centered on flaring states.
These limitations were largely imposed by observational constraints at the time, but they may nevertheless introduce sample-selection bias and capture only local or state-dependent features, thereby making it more difficult to trace the long-term evolution of color behavior.} 
With the advent of high-temporal sampling surveys, it is now possible to study the color variability of blazar populations in a more systematic and unbiased manner.~\citet{Negi.22.mn}, using two years of Zwicky Transient Facility (ZTF) survey data~\citep{Bellm.19.pasp}, conducted the first statistical analysis of color variability in approximately 900 blazars, providing new insights into the debate between ``continuity'' and ``dichotomy'' and revealing the intrinsic differences between blazar sub-types.

In this work, we extend the investigation of blazar color behavior using the larger homogeneous sample, building upon the pioneering unbiased color variability study by~\citet{Isler.17.apj}.
Our sample contains {1149} blazars spanning an observational period of approximately six years, with each source having at least 30 quasi-simultaneous g-band and r-band observations (i.e., taken within 30 minutes).
{
In this work, we aim to systematically investigate the optical color behavior of blazars on both the full six-year timescale and shorter timescales within the same monitoring baseline, and to further examine its dependence on source brightness and variability amplitude across different blazar subclasses.}

The structure of the paper is as follows: Section~\ref{sec:data} describes the data used in this study and how they were obtained; 
{Section~\ref{sec:results} analyzes the brightness and color behavior of the sources, as well as the correlation of color behavior with brightness and flux variability amplitude;}
Section~\ref{sec:dis} discusses the underlying physics behind the distributions and correlations of these statistical observables; Section~\ref{sec:con} presents the conclusions.

\section{Data and sample selection}
\label{sec:data}

\begin{figure}
    \includegraphics[width=8.5cm]{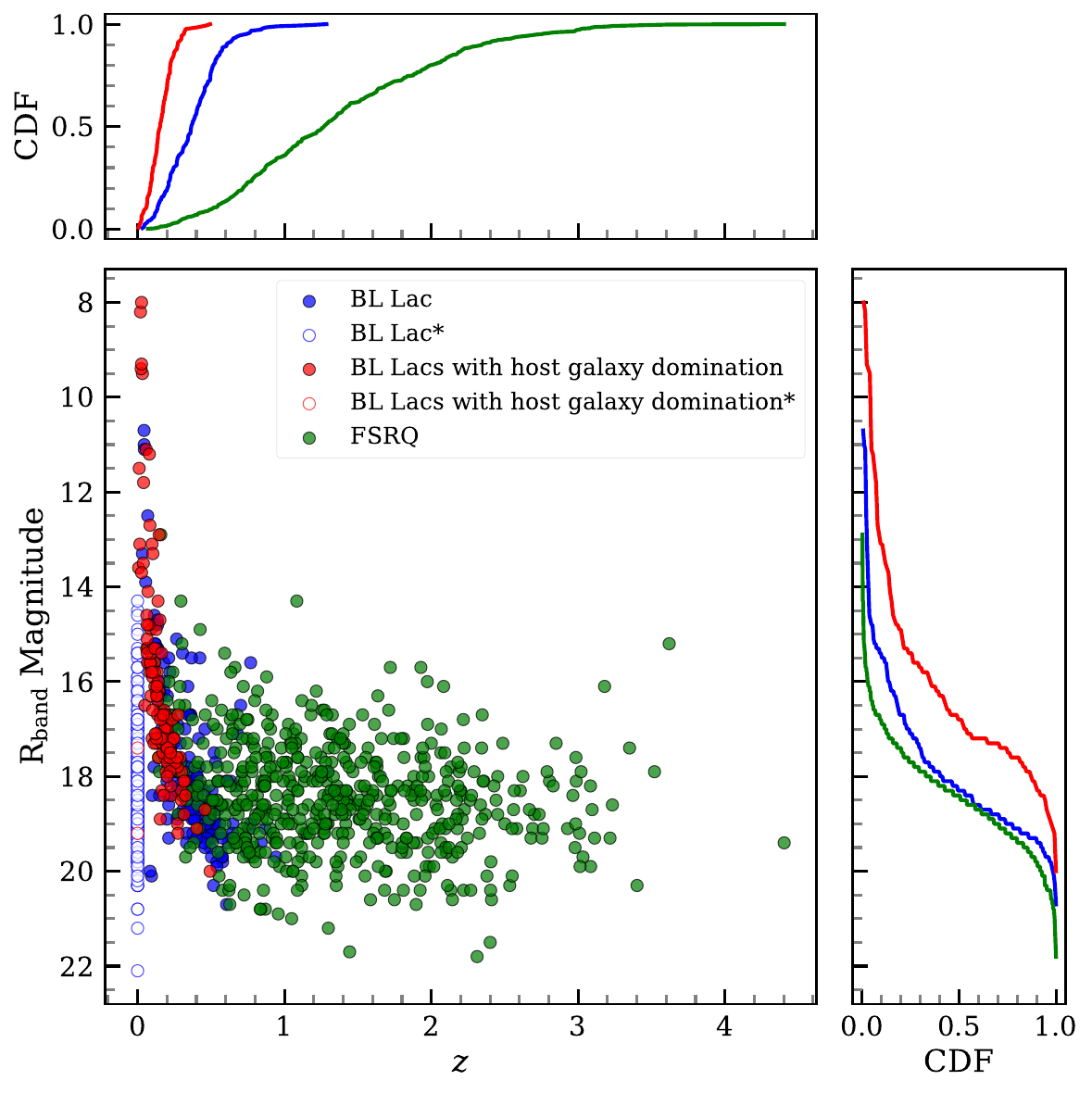}
    \caption{The redshift and magnitude distributions of the {1149} blazars in our final sample. Sources with unknown redshifts were assigned a dummy value of z=0 {and are marked with an asterisk in the legend. 
    The cumulative distribution functions (CDFs) for both redshift and magnitude are shown in the top and right panels, excluding sources with unknown redshifts.}}
    \label{fig:simple}
\end{figure}

We use the ZTF Light Curve API~\footnote{\url{https://github.com/zvanderbosch/ZTF_tools/tree/main}} to perform a cone search based on the RA-Dec coordinates of the Roma-BZCAT catalogue~\citep{Massaro.15.apss}, the most complete list of all blazars detected in multi-frequency surveys to date, containing 3561 blazars.
The default search radius (1.5 arcseconds) is small to minimize the likelihood of multiple sources falling within the search area. After retrieving the light curves, the data is separated into g-, r-, and i-band light curves. Due to inadequate sampling of i-band observations for the majority of targets, the i-band data were excluded from the present analysis to ensure analytical reliability.
Based on the g- and r-band observations, a parent sample of 2650 sources was initially constructed to serve as the basis for subsequent analyses.
To ensure data reliability, low-quality detections were excluded in accordance with the standard guidelines provided in the ZTF data release and further refined using empirical criteria established by~\citet{Negi.22.mn}.
A detailed description is provided below.
We corrected the observed magnitudes for Galactic extinction on a source-by-source and band-by-band basis using extinction values obtained from the 
NASA/IPAC Infrared Science Archive (IRSA)\footnote{\url{https://irsa.ipac.caltech.edu/frontpage}}, which are based on the recalibrated dust maps of~\citet{Schlafly.11.apj}.
To remove any outliers from the light curve, we perform sigma clipping using the \textsc{Astropy} function
\texttt{sigma\_clip}\footnote{\url{https://docs.astropy.org/en/stable/api/astropy.stats.sigma_clip.html}}
with $\sigma = 3$ for each source.
We also performed barycentric corrections to the GPS timestamps of our datasets using the Astropy~\citep{Astropy.18.aj} Python package.
{We matched quasi-simultaneous observations by minimizing $|\Delta t|$ within a 30-minute threshold, without imposing temporal ordering.
We retained only sources with at least 30 quasi-simultaneous data points for the analysis.}
{After removing 62 blazars of uncertain type, the final sample contains 1149 sources: 589 BL Lacs and 560 FSRQs.\footnote{For statistical analysis, BL Lac candidates are included in the BL Lac subsample because their observed properties are broadly consistent with those of BL Lacs.}}
{According to~\citet{Massaro.15.apss}, 127 BL Lacs in the sample} exhibit spectral energy distributions (SEDs) in which the host galaxy emission significantly dominates over the nuclear emission.
The basic information about the sample is summarized in Table~\ref{tab:source_data} and Figure~\ref{fig:simple}.

\section{analysis and results}
\label{sec:results}
\subsection{Brightness variability}

We determine the amplitude of variability ($\Psi$) and the fractional root mean square (rms) variability amplitude ($F_\mathrm{var}$) of each source using the expressions introduced by~\citet{Heridt.96.aa}
and~\citet{Vaughan.03.mn}, respectively, to quantify the brightness variability in the blazar light curves.

The $\Psi$ is defined as follows:
\begin{equation}
\Psi=\sqrt{\left(A_{\max }-A_{\min }\right)^2-2 
\sigma^2} ,
\end{equation}
where $A_{\max}$ and $A_{\min}$ are the maximum and minimum values of each light curve and $\sigma$ the measurement errors.
The $F_\mathrm{var}$ is a commonly employed metric for quantifying the variability of an astronomical object's light curve. It measures the relative amplitude of flux variations, normalized by the mean flux, while accounting for the uncertainties in the observational data.
It is defined as:
\begin{equation}
F_{\mathrm{var}}=\sqrt{\frac{S^2-\overline{\sigma_{\mathrm{err}}^2}}{\bar{x}^2}}
\end{equation}
where $S$ is the sample variance, $\sigma_{\mathrm{err}}^2$ is the mean squared error of the flux measurements, and $\bar{x}$ is the mean flux. 
The uncertainty in $F_\mathrm{var}$ is given by:
\begin{equation} 
\operatorname{err}\left(F_{\mathrm{var}}\right) 
= \sqrt{\left(\sqrt{\frac{1}{2 N}} \cdot \frac{\bar{\sigma}_{\mathrm{err}}^2}{\bar{x}^2 F_{\mathrm{var}}}\right)^2 + \left(\sqrt{\frac{\bar{\sigma}_{\mathrm{err}}^2}{N}} \cdot \frac{1}{\bar{x}}\right)^2}
\end{equation}
For further details, please refer to Appendix B of \citet{Vaughan.03.mn}'s paper.
The $\Psi$ and $F_{\mathrm{var}}$ parameters for all sources have been calculated, and the $F_{\mathrm{var}}$ distributions are shown in Figure~\ref{fig:varaibility_bllac_fsrq}.
{To avoid redundant parallel analyses in two adjacent optical bands, we use the r-band light curve as the primary tracer of optical variability in this paper. Since the main goal is to characterize the overall variability properties of the sources rather than to compare band-dependent variability in detail, a single representative band is sufficient for this purpose.}

\subsection{Color variability}
{Color serves as a proxy for the spectral slope, and changes in color reflect the underlying spectral variations. The color index--magnitude diagram therefore provides useful clues to the physical processes responsible for blazar variability. Throughout this paper, color behavior is characterized using the Pearson correlation coefficient, $r_{\rm p}$, between the color index $(r-g)$ and the $r$-band magnitude, with statistical significance assessed by the corresponding $p$-value. A source is defined as exhibiting a significant BWB trend when $r_{\rm p}\leq -0.5$ and $p<0.05$, and a significant RWB trend when $r_{\rm p}\geq 0.5$ and $p<0.05$. Cases with $|r_{\rm p}|<0.5$ or $p\geq 0.05$ are classified as showing no significant color trend.}
We first perform statistical analysis of the color behavior over the duration of the entire light curve for each source in BL Lacs and FSRQs. 
To study the temporal evolution of the color index-magnitude correlation in the photometric time series data and identify mutation points in the color evolution to reveal possible physical mechanism transitions (such as enhanced jet activity, variations in accretion rate, etc.), we used a sliding window method to determine the {short-term color index-magnitude} relationship. 
{Each sliding window contains 30 data points, with a step size of 10 data points, and only complete windows are included in the short-term color behavior analysis. This criteria provides a balance between temporal resolution and statistical robustness: a smaller step size would lead to highly redundant, strongly correlated neighboring windows, whereas a larger step size would reduce sensitivity to local state transitions. Under this scheme, no more than 9 data points are excluded for any individual source, while the statistical instability associated with insufficient sampling is effectively reduced. In each window, the short-term color behavior is classified according to the same criteria used throughout this work.}
{In each window, short-term color behavior is assessed using the same criteria adopted throughout this paper.}
We refer to this method as Sliding Window Correlation State Tracking for {Short-term} Color Behavior (\textsc{SWCST for SCB}).
{This method is particularly well suited to blazars, whose variability is intrinsically non-stationary and may involve multiple emission components as well as different phases (i.e., brighter state and fainter state).}

\subsubsection{Overall color behaviors}

We find that about {14.7\% (68/462)} of BL Lacs {with negligible host-galaxy contamination} exhibit a significant BWB trend, while only {2.3\% (11/462)} of BL Lacs exhibit a RWB trend over the duration of the entire light curve. 
Among {the 560} FSRQs, about {8.8\% (49/560)} of the sources exhibit an overall BWB trend 
over the full light curve, and about {14.1\% (79/560)}  of the sources exhibit an overall RWB trend.
{Furthermore, among the 127 BL Lacs significantly affected by host-galaxy contamination, only 5 show a BWB trend, whereas 28 exhibit an RWB trend.}
The upper panels of Figures~\ref{fig:case_bllac} and~\ref{fig:case_fsrq} show representative examples of BWB and RWB trends from BL Lacs and FSRQs, respectively.
These results indicate that the BWB trend is more prevalent in BL Lacs, while FSRQs tend to exhibit a RWB trend.

\subsubsection{Short-term color behavior}

\begin{figure*}
    \includegraphics[width=8.5cm]{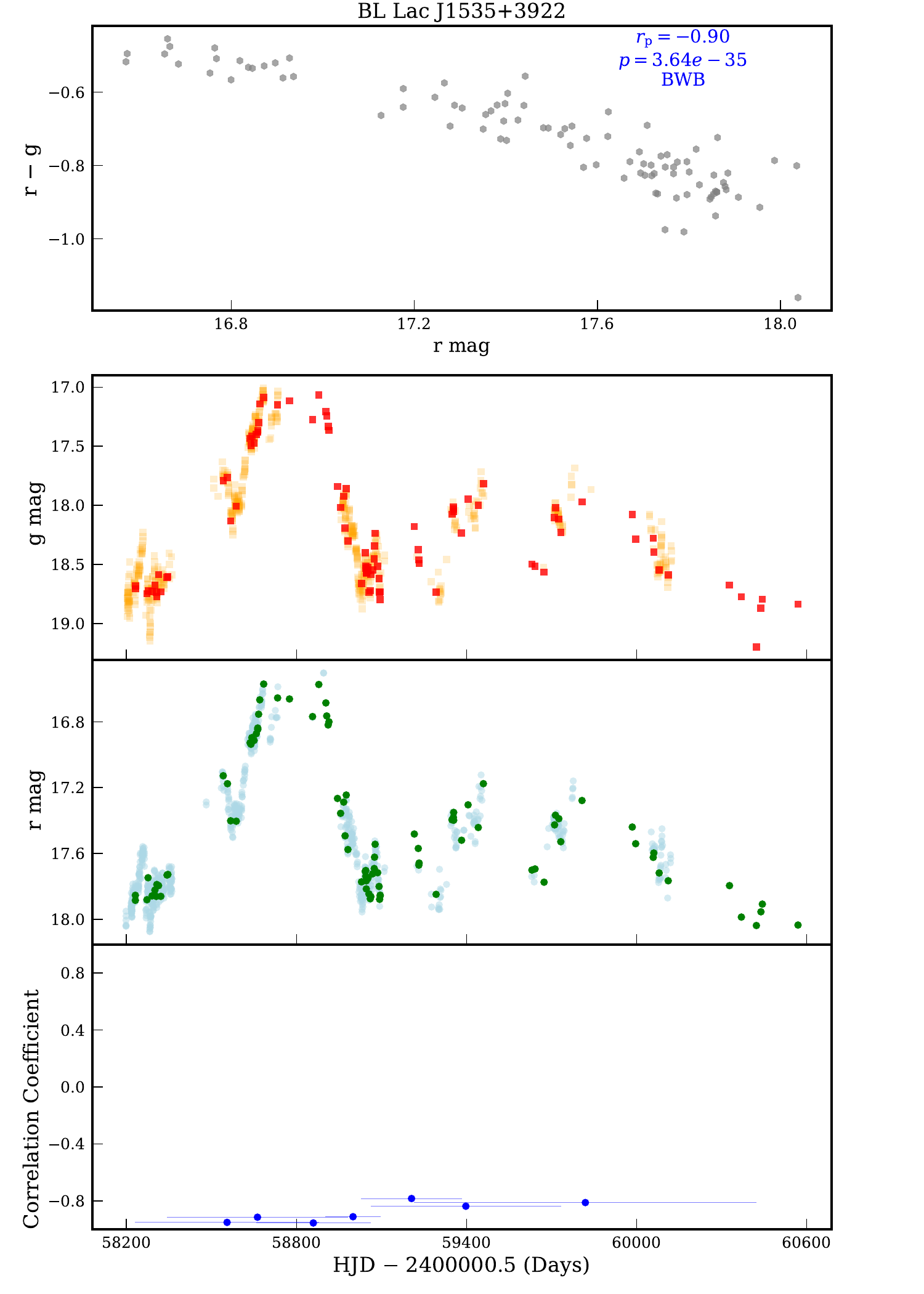}
    \includegraphics[width=8.5cm]{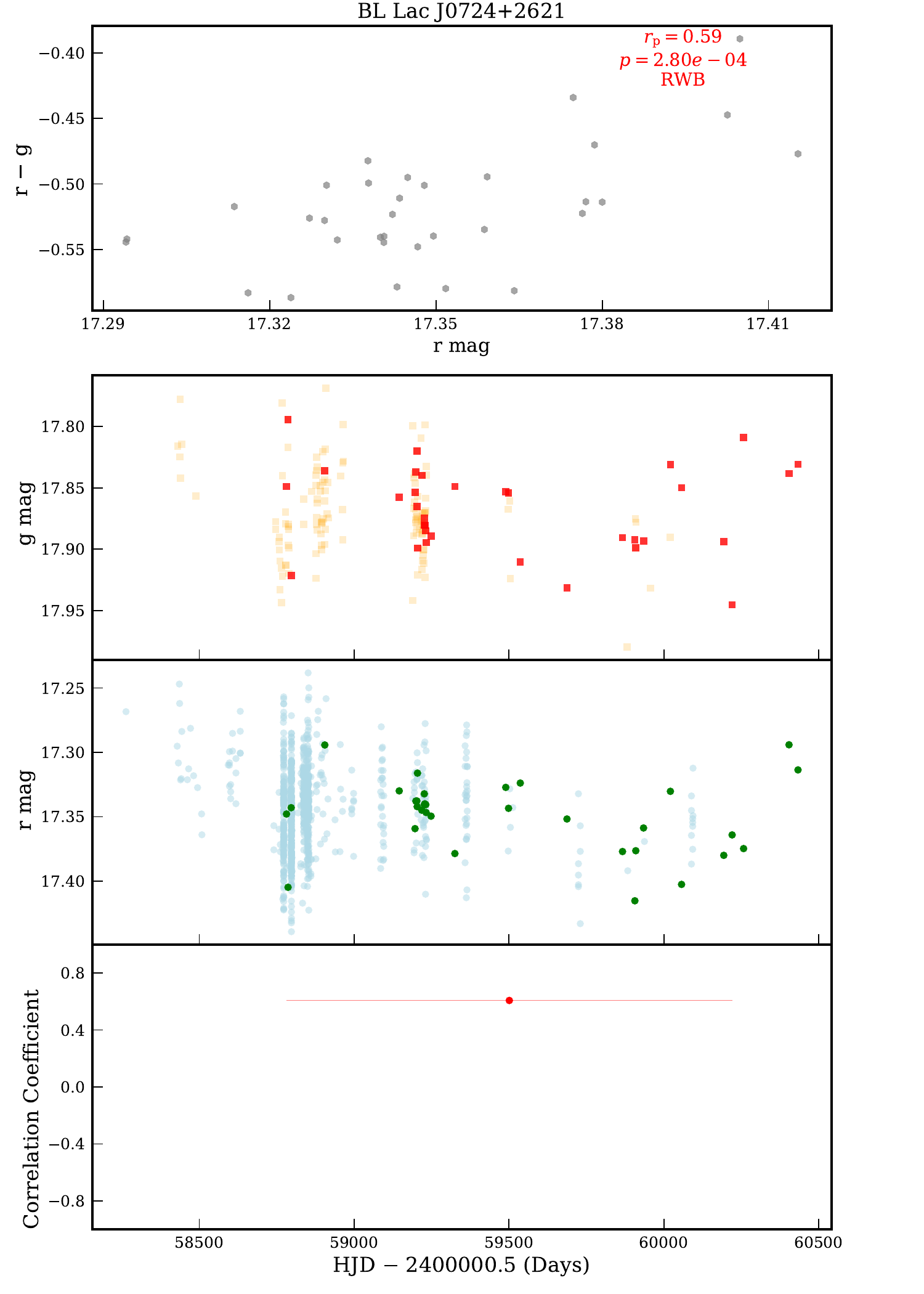}
    \caption{
    Color index $(\mathrm{r} - \mathrm{g})$ vs. r-band magnitude, light curve of ZTF data, and short-term color behavior variation of two BL Lac cases. In each case, the upper panel shows the color index $(\mathrm{r} - \mathrm{g})$ vs. r-band magnitude; {the upper-right annotation in the upper panel indicates the Pearson correlation coefficient ($r_p$), p-value ($p$), and long-term color behavior.} The two middle panels display the g-band and r-band light curves, with dark-colored dots indicating quasi-simultaneous data points, which means the two most recent g-band and r-band observations within a 30-minute interval. The bottom panel shows the variation in the short-term correlation coefficient between color index $(\mathrm{r} - \mathrm{g})$ and r-band magnitude over time, with colored dots representing the median values for each time window, and the length of the semi-transparent bars corresponding to the full duration of each time window; blue and red correspond to the BWB and RWB trends, respectively.}
    \label{fig:case_bllac}
\end{figure*}

\begin{figure*}
    \includegraphics[width=8.5cm]{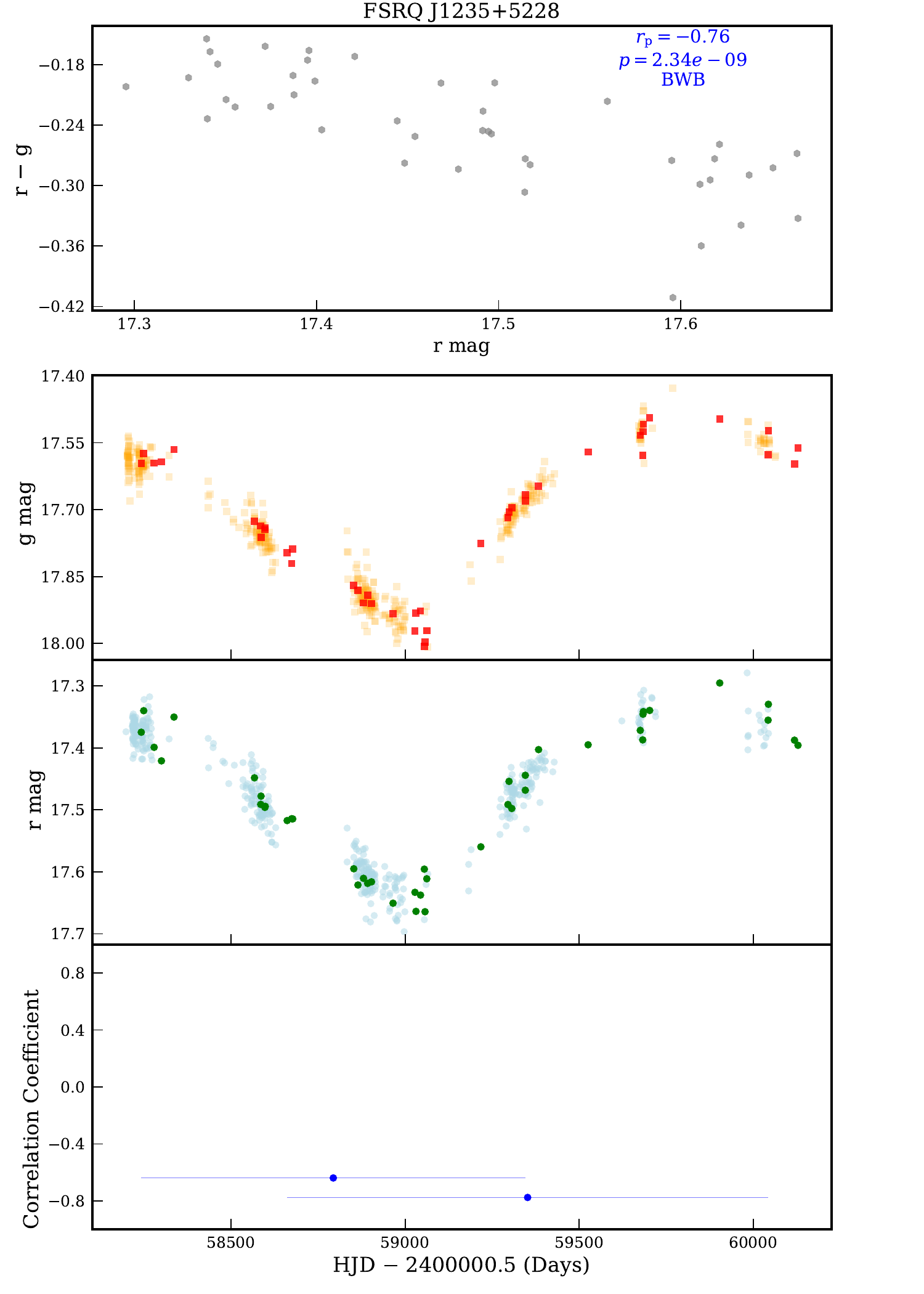}
    \includegraphics[width=8.5cm]{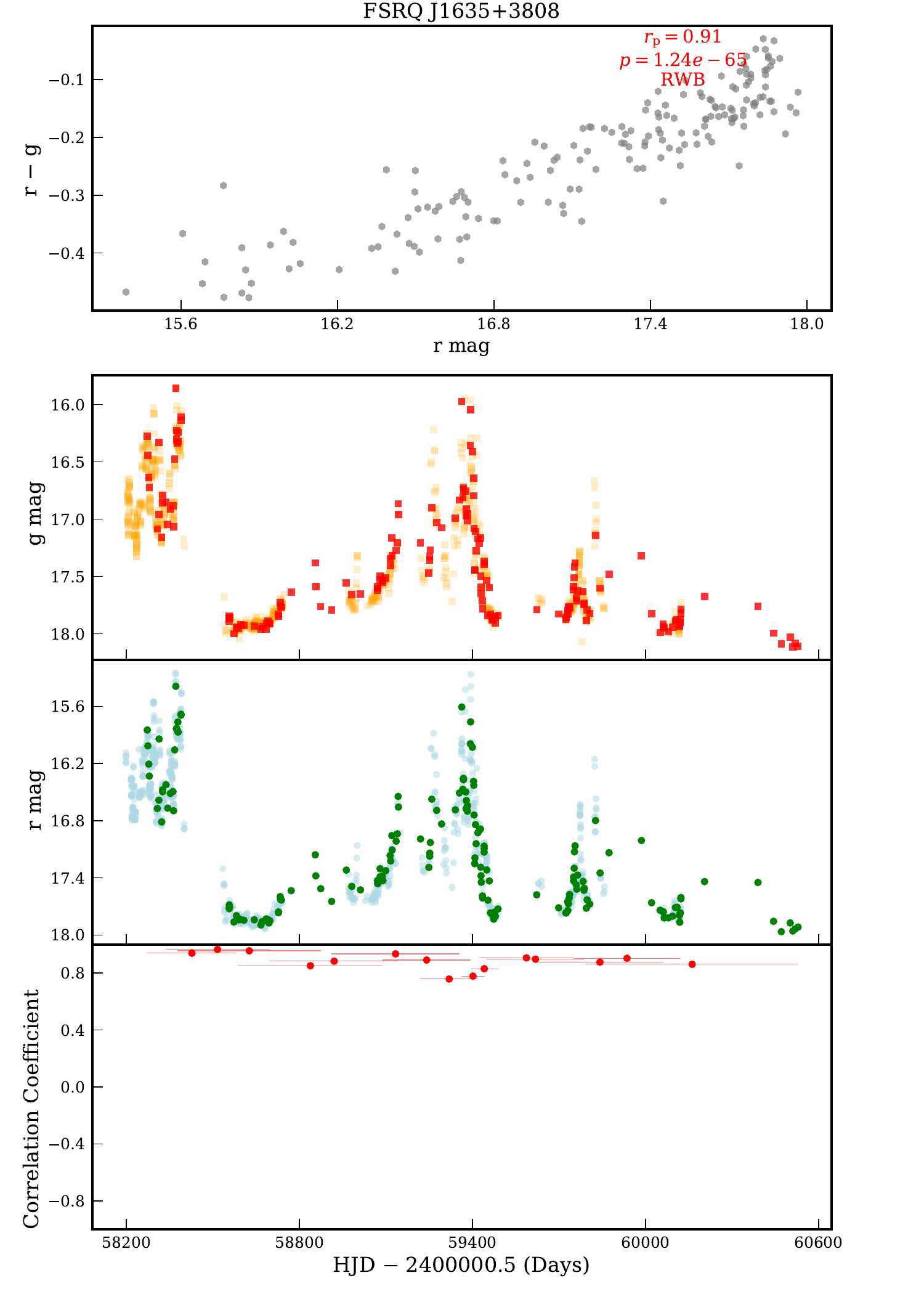}
    \caption{Color index $(\mathrm{r} - \mathrm{g})$ vs. r-band magnitude, light curve of ZTF data, and short-term color behavior variation of two FSRQ cases. 
    The elements in this figure are consistent with those in Figure~\ref{fig:case_bllac}.}
    \label{fig:case_fsrq}
\end{figure*}

We used the \textsc{SWCST for SCB} method to investigate {short-term} color behaviors within certain time intervals during the observation period (i.e., the six-year data released by DR22). 
The bottom panels of Figures~\ref{fig:case_bllac} and~\ref{fig:case_fsrq} show the evolution of the color index-magnitude correlation.
{This color behavior evolution may reflect short-term changes in the physical conditions of the source, leading to variations in both color (specturm) and brightness.}

\begin{figure*}
    \includegraphics[width=18cm]{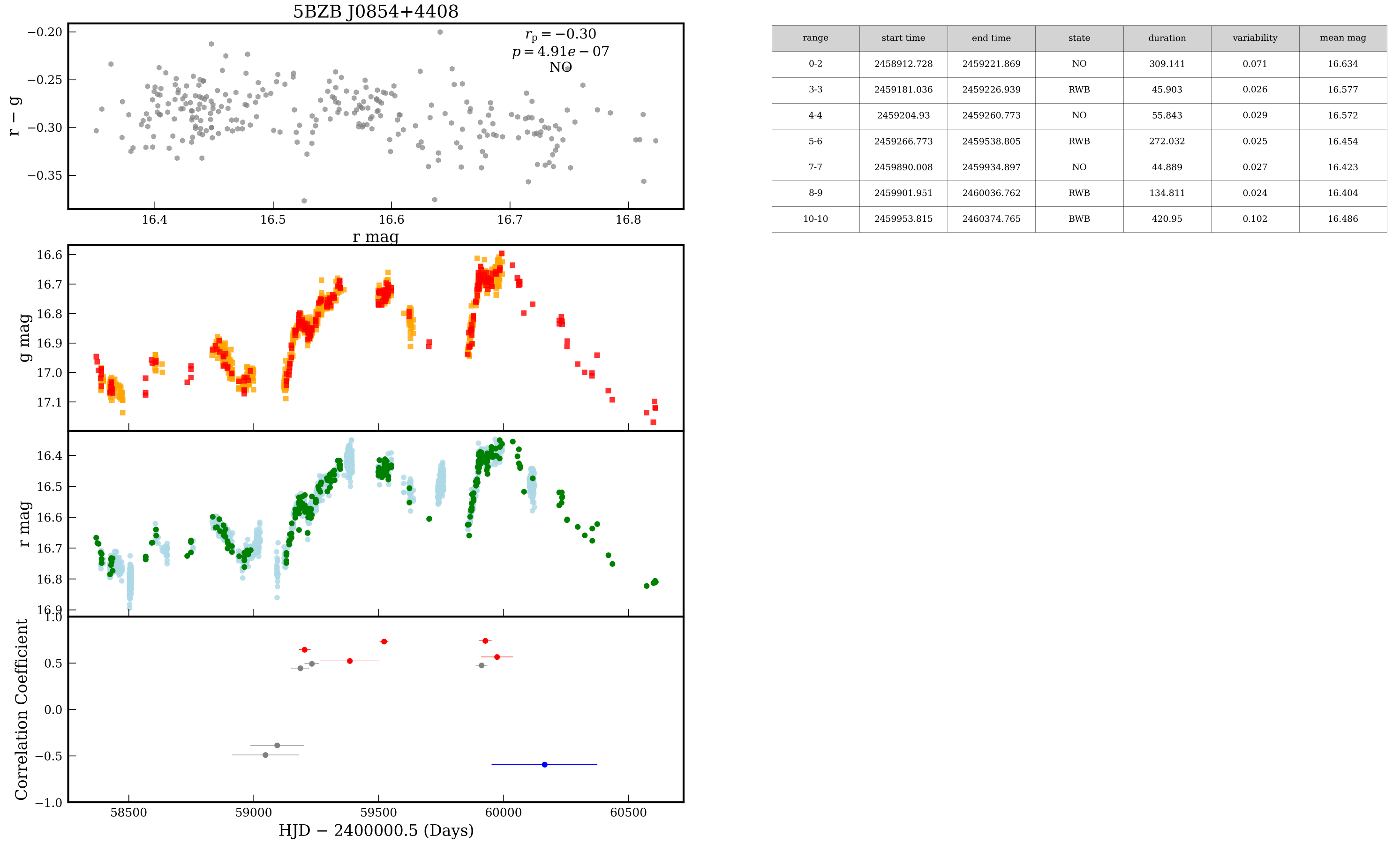}
    \caption{Color index (r$-$g) versus r-band magnitude, ZTF light curves, and short-term color behavior variations for the BL Lac 5BZB J0854+4408, which shows no significant long-term color trend but exhibits short-term BWB and RWB episodes over the six-year monitoring period. The elements in the left panels are the same as those in Figure~\ref{fig:case_bllac}; the gray symbols in the bottom panel indicate intervals with no significant color trend. The table summarizes the start and end times, color states, durations, variability amplitudes, and mean magnitudes of the identified intervals. Statistically insignificant results with $p > 0.05$ were removed from the analysis and therefore excluded from the tables and figures.
    Figures for all sources are publicly available on Zenodo at \dataset[DOI: 10.5281/zenodo.19812075]{https://doi.org/10.5281/zenodo.19812075}.}
    \label{fig:case_no_trend}
\end{figure*}

In addition, among those sources that do not show a color behavior globally, some sources show a {short-term} color behavior. 
Of the {388} BL Lacs that exhibited no color trend in their long-term light curves, {48} sources showed short-term BWB trends within certain time intervals, while {40} sources exhibited short-term RWB trends. An additional {5} sources displayed complex variations in color behaviors, intermittently showing BWB, RWB, or no distinct color behavior.
For FSRQs, of the {432} sources that exhibited no color trend throughout the entire observation period, {12} sources showed short-term BWB trends during certain time intervals, {145} exhibited short-term RWB trends, and another {5} sources demonstrated complex behavior across different time intervals. 
Figure~\ref{fig:case_no_trend} presents an example of complex short-term color behaviors for an individual target, which shows no overall color behavior in its light curves.
All other cases are provided in the supplementary materials.

\begin{table}[ht]
\begin{flushleft}  
\scriptsize
\caption{The essential parameters of the blazar sample examined in this study.}
\begin{tabular}{cccccc}
\hline
\textbf{Source Name} & \textbf{RA} & \textbf{Dec} & \textbf{Type} & \textbf{$r_{\text{mag}}$} & \textbf{$z$} \\
(1) & (2) & (3) & (4) & (5) & (6) \\
\hline
J0007$+$4712 & 00 07 59.97 & $+$47 12 07.7  & BL Lac  & 18.2 & 0.28 ? \\
J0035$+$1515 & 00 35 14.7  & $+$15 15 04.21 & BL Lac  & 16.8 & ?? \\
J0015$+$3536 & 00 15 27.96 & $+$35 36 39.59 & BL Lac  & 19.8 & 0 \\
J0018$+$2947 & 00 18 27.78 & $+$29 47 30.8  & BL Lac  & 18.4 & 0.1 ? \\
J0800$+$1645 & 08 00 18.78 & $+$16 45 56.98 & BL Lac Can  & 18.9 & 0.20 ? \\
J1123$+$7229 & 11 23 49.16 & $+$72 29 59.6  & BL Lac Can  & 18.8 & ?? \\
J0048$+$3157 & 00 48 47.14 & $+$31 57 25.2  & BL Lac* & 13.1 & 0.015 \\
J0103$+$1526 & 01 03 26.01 & $+$15 26 24.79 & BL Lac* & 17.2 & 0.246 \\
J0004$+$4615 & 00 04 16.17 & $+$46 15 18.21	& FSRQ    & 20.4 & 1.81 \\
J0005$+$3820 & 00 05 57.18 & $+$38 20 15.21	& FSRQ    & 17.6 & 0.229 \\
J0006$+$2422 & 00 06 48.78 & $+$24 22 36.51 & FSRQ    & 18.8 & 1.684 \\
J0012$+$3353 & 00 12 47.35 & $+$33 53 38.61 & FSRQ    & 20.4 & 1.682 \\
J0048$+$3157 & 00 48 47.14 & $+$31 57 25.2  & BL Lac* & 13.1 & 0.015 \\
J0059$-$0150 & 00 59 16.91 & $-$01 50 17.69 & BL Lac* & 17.1 & 0.114 \\
-- & -- & -- & -- & -- & -- \\
\hline
\label{fig:info_table}
\end{tabular}
\smallskip  
\textbf{Note:}  
Column (1): Source Name. 
Column (2): RA. 
Column (3): Dec.
Column (4): Type. The ``*'' indicates that, based on the SED fitting results, the radiation from the host galaxy of this source is significantly stronger than that from its nuclear region. 
The label "BL Lac cand" indicates that this source is a BL Lac candidate.
Column (5): $r$-band magnitude from the Roma-BZCAT catalogue;
Column (6): Redshift $z$.
The "?" indicates uncertainty in the redshift measurement;
The "??" indicates sources with unknown redshifts.
(This table is available in its entirety in machine-readable form.)

\label{tab:source_data}

\end{flushleft}
\end{table}

\begin{figure}
\includegraphics[width=8.5cm]{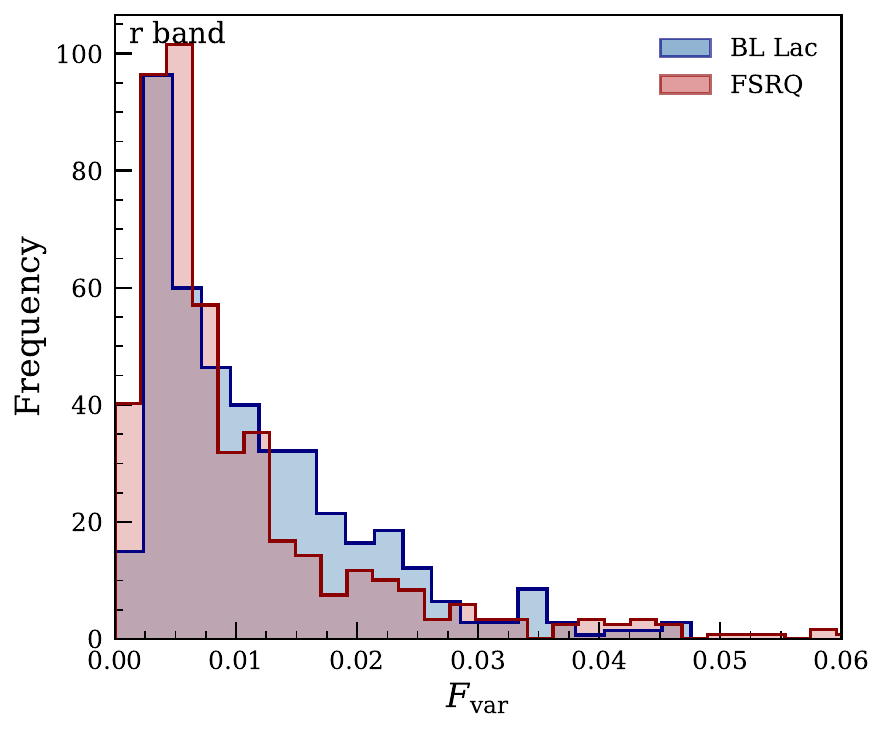}
    \caption{
    Histograms of quantities characterizing brightness variability for different types of blazars.
    Each steelblue and indianred histogram represents the distribution of brightness variability for BL Lacs and FSRQs, respectively.}
    \label{fig:varaibility_bllac_fsrq}
\end{figure}

\subsection{Correlation of color behavior with brightness and variability}

\begin{figure*}
\includegraphics[width=9cm]
    {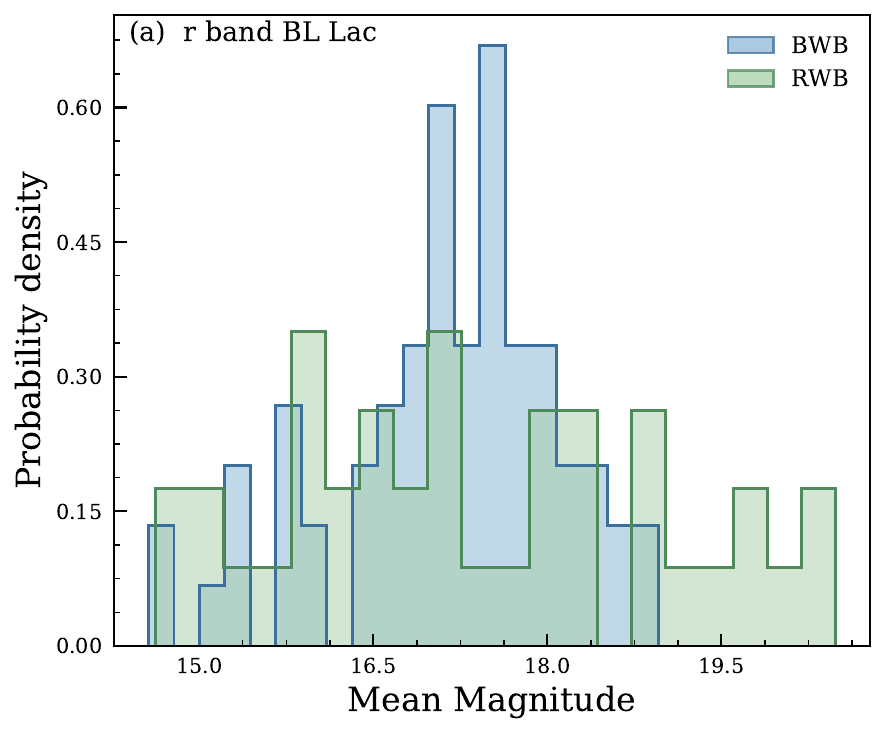}
    \includegraphics[width=9cm]
    {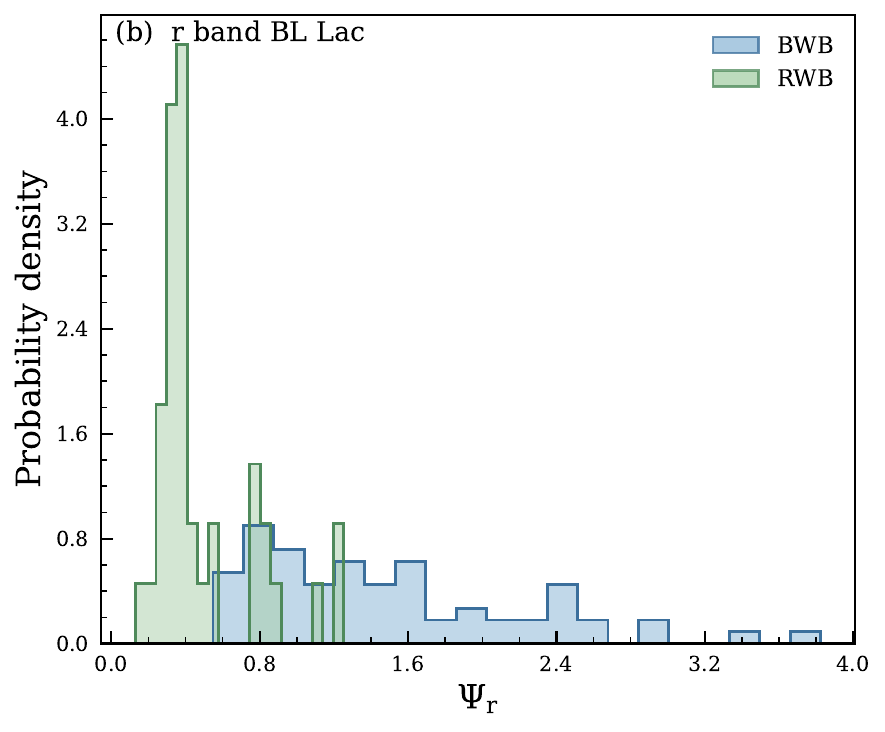}
    \includegraphics[width=9cm]{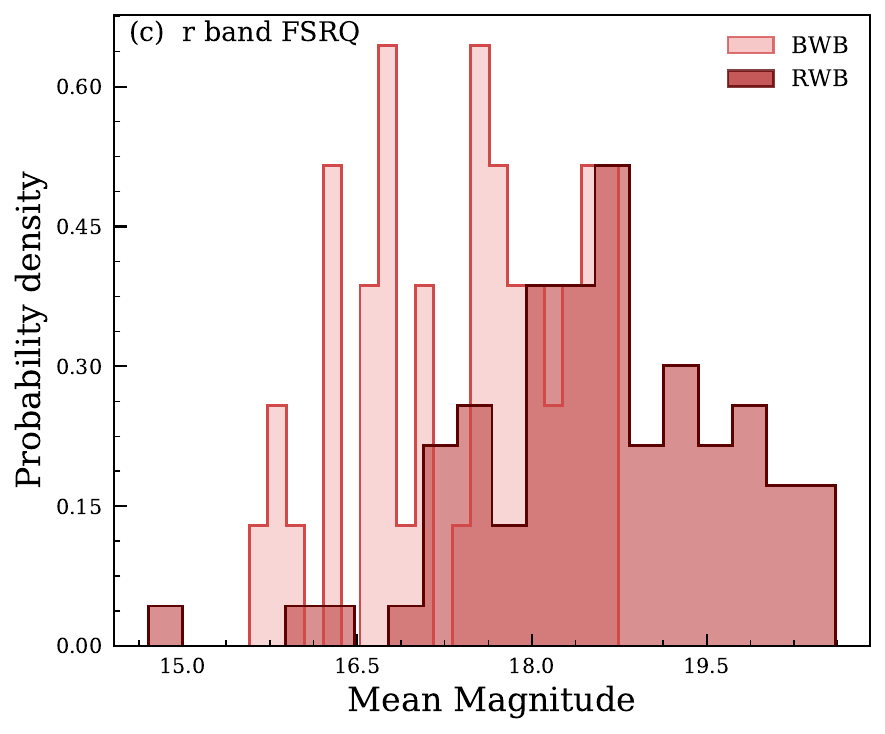}
    \includegraphics[width=9cm]
    {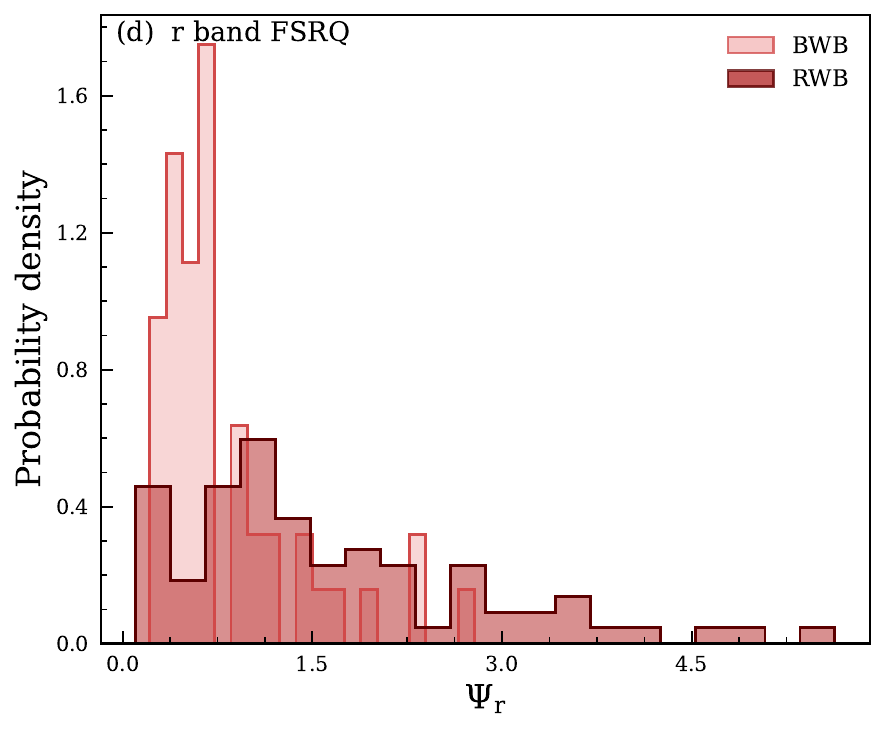}
    \caption{Probability density distributions of the mean flux and variability amplitude in the r band for BL Lacs and FSRQs. Panels (a) and (c) show the distributions of the mean flux for BL Lacs and FSRQs, respectively, while Panels (b) and (d) present the corresponding distributions of the variability amplitude $\Psi_{r}$.}
    \label{fig:brightness_variability}
\end{figure*}

\subsubsection{Overall Statistical Correlations}
{
Using a six-year dataset, we revisit the statistical connection between blazar color behavior, brightness state, and variability. The longer temporal baseline allows us to better trace long-term behaviors that may be missed in shorter-term datasets and to derive more robust statistical distributions for different subclasses.
As shown in Figure~\ref{fig:brightness_variability}(a) and~\ref{fig:brightness_variability}(c), for both BL Lacs and FSRQs, sources with higher mean brightness over the six-year monitoring period are more likely to exhibit a BWB trend, whereas sources with lower mean brightness are more likely to show an RWB trend. For FSRQs, this statistical tendency is broadly consistent with the results reported by~\citet{Ikejiri.11.pasp,Mao.16.apss}.
However, our results do not support the idea that the BWB trend is universal among blazars. 
We suggest that the observed BWB or RWB trend of a source is more likely controlled by the relative contributions of different radiative components at different brightness levels.
}

{We also analyzed the correlation between the two color behaviors and the variation amplitude.
These distributions (Figure~\ref{fig:brightness_variability}(b) and~\ref{fig:brightness_variability}(d)) show that the relationship between color behavior and variability amplitude differs between BL Lacs and FSRQs. For BL Lacs, the BWB subsample extends to systematically larger values than the RWB subsample, suggesting that stronger variability is more commonly associated with the BWB trend. In contrast, for FSRQs, the RWB subsample exhibits a broader distribution and a more pronounced high-$\Psi$ tail, indicating that larger variability amplitudes are preferentially linked to the RWB state.
}

\subsubsection{Short-term Statistical Correlations}
{It is also important to note that the “brighter” and “fainter” phases discussed in previous studies~\citep{Ikejiri.11.pasp,Mao.16.apss,Negi.22.mn} were defined based on the overall source brightness during the monitoring period, rather than by strictly tracing the temporal transition between bright and faint states within an individual source. 
Given that blazars typically undergo strong and frequent flux variations over long-term monitoring, defining bright and faint phases solely in terms of the long-term mean brightness is insufficient to fully capture the correspondence between color behavior and short-term brightness states.
Based on this consideration, we further analyzed the relationship between the average magnitude of each source and the color trend within a short-term time window.}

\begin{figure}
    \includegraphics[width=8.5cm]{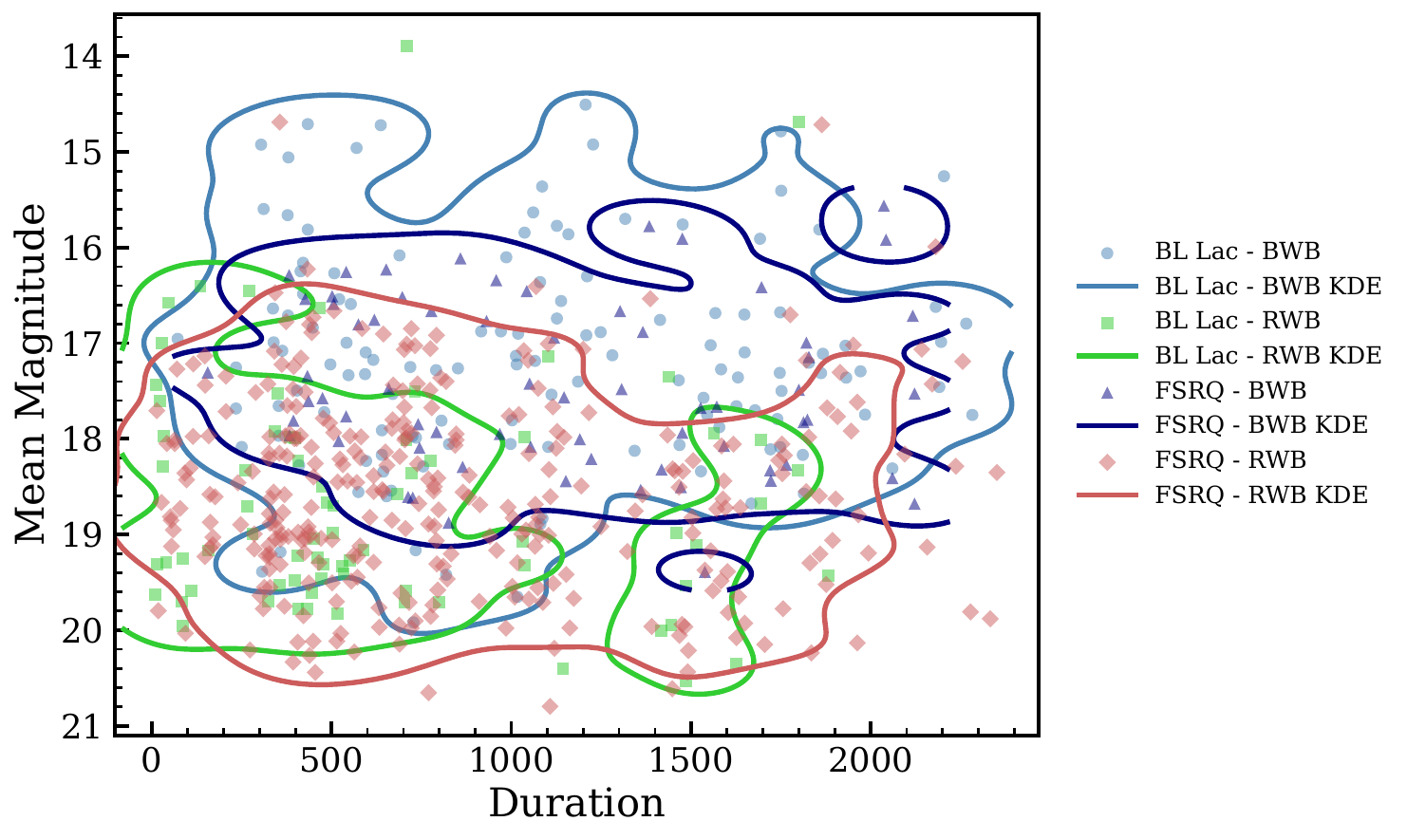}
    \caption{Time-window interval versus mean magnitude for the BL Lacs showing the BWB trend, BL Lacs showing the RWB trend, FSRQs showing the BWB trend, and FSRQs showing the RWB trend. The symbols denote individual data points, while the solid curves indicate the low-density contours derived from two-dimensional Gaussian KDE, tracing the overall extent of each distribution in the parameter space. The contour level corresponds to 12\% of the maximum KDE value for each subgroup.}
    \label{fig:short_term_brightness}
\end{figure}

\begin{figure}
    \includegraphics[width=8.5cm]{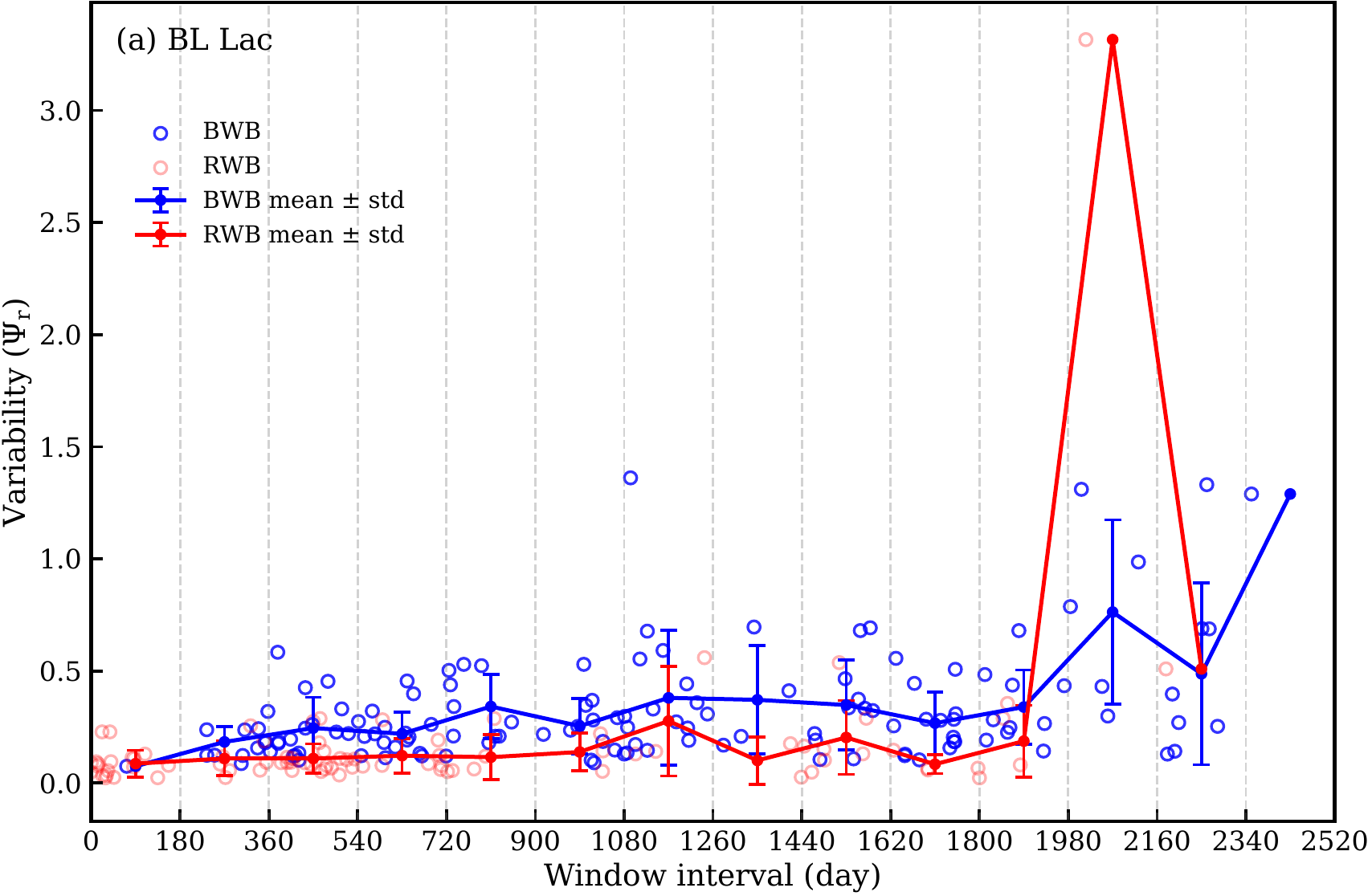}
    \includegraphics[width=8.5cm]{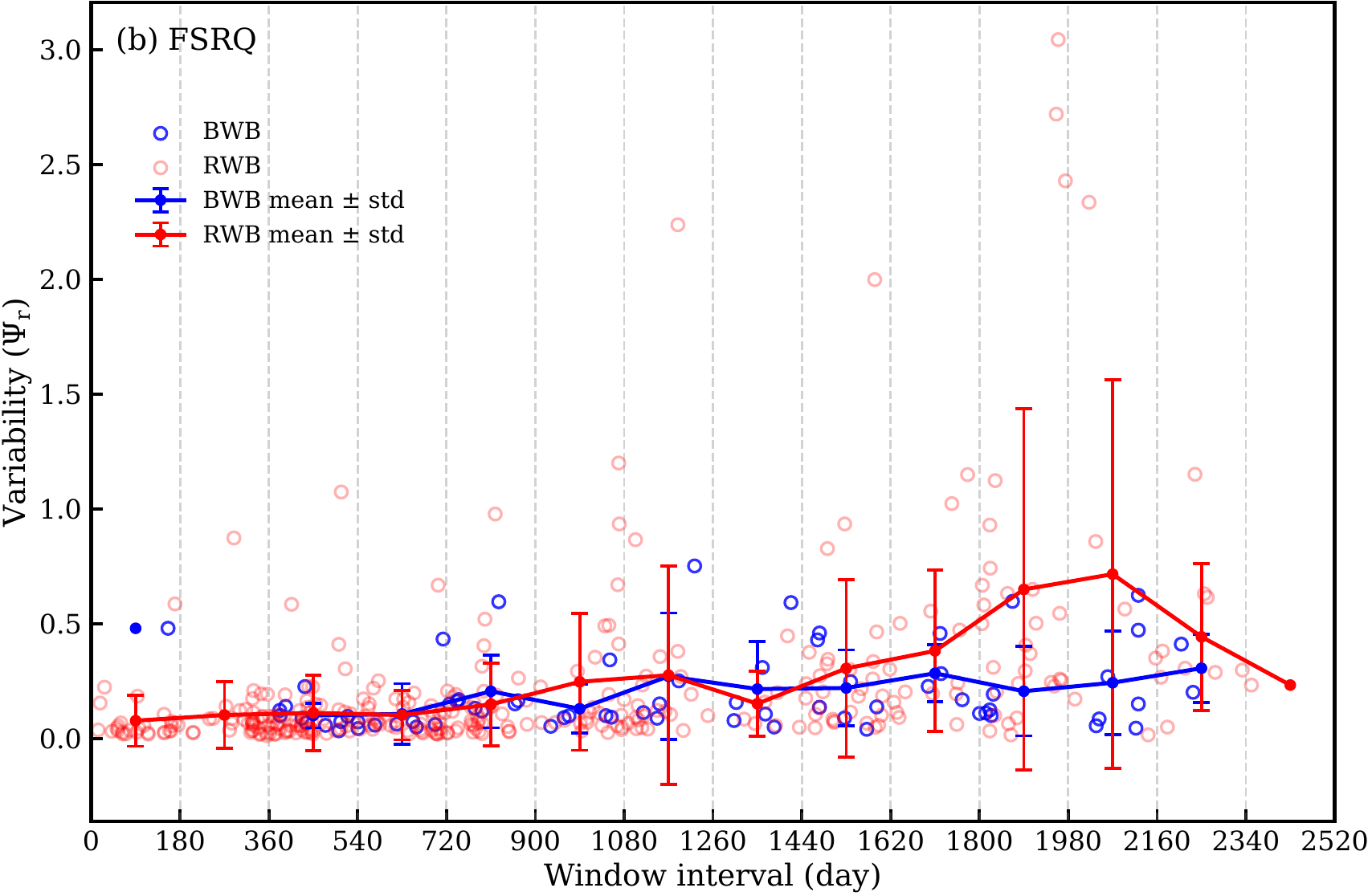}
    \caption{Variability amplitude in the r band, as a function of time-window interval for the BWB and RWB subsamples of blazars. 
    Panel (a) shows the BL Lac sample, and Panel (b) shows the FSRQ sample. Open circles represent individual measurements in different time windows, while the solid points connected by lines denote the mean $\Psi$ in each time-window bin (fixed at 180 days); the error bars indicate the corresponding standard deviation. The prominent RWB peak in the 1980–2160 day bin of the Panel (b) is based on only one source and is therefore likely affected by small-number statistics.}
    \label{fig:short_term_variability}
\end{figure}

{Figure~\ref{fig:short_term_brightness} compares the distributions of the BWB and RWB subsamples in the duration-mean magnitude plane for BL Lacs and FSRQs, with the overall distributions traced by overlaid two-dimensional kernel density estimation (KDE) contours.
The short-term time-window analysis yields the same statistical correlation: in both BL Lacs and FSRQs, sources in a brighter state are more likely to exhibit a BWB trend, whereas sources in a fainter state are more likely to show an RWB trend.
Figure~\ref{fig:short_term_variability} shows the dependence of the r-band variability amplitude, $\Psi_r$, on the time window interval for the BWB and RWB subsamples of BL Lacs and FSRQs. In both subclasses, the variability amplitude generally increases with increasing time window interval, indicating that larger amplitudes are more readily revealed on longer timescales. However, the subclass dependence is markedly different. 
For BL Lacs, the BWB subsample generally shows larger mean than the RWB subsample over most of the sampled timescales. 
In contrast, for FSRQs, the RWB subsample tends to exhibit systematically larger mean 
values and a broader scatter than the BWB subsample, especially at long window intervals. These results suggest that the statistical connection between color behavior and variability amplitude is strongly subclass dependent, consistent with the overall statistical results in section 3.3.1.
}

\section{Discussion}
\label{sec:dis}

\subsection{Examining the physical processes responsible for the distribution of brightness variability}
As shown in Figure~\ref{fig:varaibility_bllac_fsrq}, when  $F_\mathrm{var} < 0.035$, the fractional rms variability amplitude ($F_\mathrm{var}$) distribution of BL Lacs is consistently greater than that of FSRQs. However, across the entire sample, FSRQs exhibit a small subset of sources with significantly higher variability than BL Lacs.
In general, variations in the viewing angle of the emitting knots or jet regions~\citep{Raiteri.17.na, Britzen.18.mn}, along with changes in the mass accretion rate~\citep{Pereyra.06.apj,Lishuangliang.08.mnras,Kokubo.15.mnras} or the fluctuations in the UV–optical region of the accretion disk (e.g., turbulence driven by magnetorotational instability~\citep{Balbus.91.apj} or convection due to increased opacity in the ultraviolet optical area of the disk~\citep{Jiangyanfei.20.apj}), can provide plausible explanations for flaring events or increases in flux.
The former is related to the relativistic jet, where even small wobbles or changes in the viewing angle can cause significant and rapid variation in the observed flux. In contrast, the latter two are associated with the accretion disk, which evolves much more slowly, contributing to the long-term variability observed in AGNs.
Our statistical results suggest that BL Lacs are more likely to be dominated by beaming effects, which correspond to the rapid flux variability, whereas FSRQs tend to be influenced by changes in the accretion disk, which correspond to the long-term flux variability.

A plausible explanation for the higher variability observed in a small subset of FSRQs is that these sources may be jet-dominated, with stronger Doppler factors and smaller viewing angles. In contrast, jet-dominated BL Lacs have weaker Doppler factors and larger viewing angles~\citep{Hovatta.09.aa, Potter.15.mn}. 
When the smaller viewing angle, associated with higher Doppler factors, experiences slight variations, the beaming effect can significantly amplify these changes, resulting in the higher variability observed in these FSRQs.
As a result, FSRQs in this sample exhibit greater variability than BL Lacs.

\subsection{Understanding the Causes of Overall and the Short-term Color Behavior Phenomena}

\subsubsection{Explanations of the BWB trend}
{Several mechanisms have been proposed to explain the BWB trend. One possible explanation is provided by the shock-in-jet scenario~\citep{Marscher.95.apj}, in which electrons are accelerated to higher energies at the shock front. Since higher-energy electrons emit higher-frequency synchrotron radiation and have shorter radiative cooling times than lower-energy electrons, the high-frequency bands are expected to exhibit stronger variability, thereby producing a BWB trend during flux enhancements.
The BWB trend may also be associated with the injection of fresh electrons triggered by internal shocks produced by collisions between relativistic shells in the jet. Such a process can increase the number of high-energy electrons in the emitting region and shift the synchrotron peak frequency ($v_\mathrm{syn}^\mathrm{peak}$) toward higher energies, further enhancing the blue end of the spectrum during bright states~\citep{Spada.01.mn, Guetta.04.aa}.}
{Variations in the Doppler factor may also contribute to the observed BWB trend. 
An increase in the Doppler factor $\delta$ simultaneously amplifies the observed flux ($F_{\nu} \propto \delta^{p}$) and shifts the observed emission toward higher frequencies ($\nu \propto \delta$)~\citep{Larionov.10.aa}.
For a convex spectrum, an increase in $\delta$ during the rising phase of a flare shifts the optical emission toward a harder, bluer part of the spectrum, producing a BWB trend; although the g- and r-band fluxes vary simultaneously, the g-band flux rises more rapidly than the r-band flux~\citep{Villata.04.aa,Papadakis.07.aa}.
If the spectrum remains a strict power law, changes in $\delta$ will brighten the source nearly achromatically, and the color index will remain essentially unchanged.
}

\subsubsection{Explanations of the RWB trend}

{
Compared to the BWB trend, the physical interpretation of the RWB trend is generally more straightforward, and is particularly easier to understand in the context of FSRQs. 
When the jet is in a relatively quiescent state, the thermal radiation component from the accretion disk tends to manifest more prominently in the optical-to-ultraviolet band~\citep{Pian.05.mn}. 
Under the standard thin disk model, the accretion disk spectrum can be approximated as multi-temperature blackbody radiation, with its characteristic frequency jointly determined by the black hole mass and the accretion rate~\citep{Shakura.73.aa, Frank.02.apa}. 
FSRQs generally host massive black holes (e.g., the sample average reported by~\citet{Zhang.24.mn} is $10^{8.78 \pm 0.30} \mathrm{M_\odot}$, with the majority of sources above $10^8 \mathrm{M_\odot}$), the peak of their thermal radiation typically resides in the ultraviolet or even the extreme ultraviolet band. Consequently, what is detected in the optical band corresponds to the rising portion of this thermal spectrum to the left of its peak (in the current work, the contribution in the g band is stronger than that in the r band), thereby manifesting as a relatively blue optical spectral shape. In contrast, the jet's synchrotron radiation in the optical band typically corresponds to a continuum component where the contribution from the lower-frequency end is stronger, thereby manifesting as a relatively red optical spectral shape. As the jet's synchrotron radiation intensifies, the contribution of this relatively red, non-thermal component to the total radiation steadily increases; this not only causes the source to brighten overall but also shifts the integrated color toward the red end of the spectrum, ultimately manifesting as the RWB trend.}

\subsubsection{A Unified Two-Component Framework}

{Although the BWB and RWB trends can be interpreted in terms of different specific physical processes, both behaviors are sometimes discussed within a more general two-component picture, since photometric and color information alone usually do not allow a reliable attribution of their physical origin.
Within this descriptive two-component picture, different color behavior may be reflected in changes in the relative contributions of two emission components with different spectral shapes and variability properties.
For example, a BWB trend may be interpreted as the increasing dominance of a variable component with a flatter spectrum over a relatively stable component with a steeper spectrum, causing the total optical spectrum to become bluer as the source brightens \citep[e.g.,][]{Gu.11.aa}.
In contrast, if the flux increase is dominated by a redder component with a steeper spectrum, or if the relative contribution of the bluer component decreases, an RWB trend is more likely to appear.
In this sense, the framework proposed by \citet{Isler.17.apj} can be regarded as a more comprehensive physical interpretation of the same general two-component picture, in which the observed color behavior depends on whether the emission is disk-dominated, jet-dominated, or in a transition state between the two.}

\subsubsection{Type- and Time-dependent Color Behavior in Blazars}
{
Based on our analysis of a six-year observational dataset, we find that BL Lacs without significant host-galaxy contamination tend to exhibit a BWB trend on the whole, whereas FSRQs more commonly show an RWB trend. 
However, the fractions of sources with significant color trends are low in both subclasses: only about 11\% of BL Lacs display a BWB trend, while about 14.1\% of FSRQs show an RWB trend. These fractions are significantly lower than those reported by \citet{Negi.22.mn} based on a two-year observational dataset, namely, about 18.5\% of BL Lacs exhibiting BWB and about 17.6\% of FSRQs exhibiting RWB. This result suggests that, over a longer temporal baseline, a single global color index-magnitude correlation is generally insufficient to characterize the color behavior of most blazars as a stable, monotonic, and long-lasting BWB or RWB pattern. 
To some extent, this reflects the fact that, as the time baseline increases, short-term flux variations and their associated color evolution are more easily averaged out and diluted in the long-term statistics, thereby reducing the significance of the overall trend. In other words, a given blazar is unlikely to remain in the same mode of spectral evolution throughout a long monitoring period. Instead, its radiative evolution may be jointly affected at different epochs by relativistic beaming associated with plasma motion within the jet, changes in the physical conditions of the jet, such as magnetic field strength and particle density, as well as shock-in-jet processes.}

{
\citet{Zhangbingkai.14.raa} subdivided complex color behaviors into five categories because they had already recognized that the simple BWB/RWB dichotomy is insufficient to encompass the full range of color evolution observed in blazars. 
\citet{Isler.17.apj} characterized the long-term optical/near-infrared color variability of 3C~279 and explained its complex behavior in terms of a unified jet-disk framework with time-dependent relative contributions from the two components.
Our results further support and extend this view from a statistical perspective, showing that such complexity is not confined to a few special cases, but is instead a common feature of long-term blazar variability.}

{It should also be noted that BL Lacs significantly affected by host-galaxy contamination are more likely to exhibit an RWB trend, with the fraction of RWB sources being markedly higher than that of BWB sources (22.0\% versus 3.9\%). The host galaxies of BL Lacs are typically evolved giant ellipticals~\citep{Urry.2000.apj,Safna.20.mn}, whose stellar populations are dominated by red giant and asymptotic giant branch (AGB) stars, with only a minor contribution from hot young blue stars; consequently, the host-galaxy emission is generally redder. This redder component may dominate over, and thus dilute or obscure, the intrinsic BWB trend associated with jet emission or Doppler boosting.}

{
Moreover, we use the \textsc{SWCST for SCB} method, trace the evolution of the color--magnitude correlation over different time intervals and thereby identify short-term color trends. 
This further suggests that the color behavior of blazars is not fixed over long timescales, but may shift between different modes, possibly in response to variations in jet conditions, shock activity, relativistic beaming, and the relative contributions of different emission components.
We present both the overall and short-term color trends for all sources in the supplementary material. These results provide a useful statistical sample for identifying sources with particularly complex spectral evolution, and thus lay the groundwork for future source-by-source investigations.}

\subsection{Exploring the Factors Influencing the Correlation Between Color Behavior and Variability}

{Previous studies have paid considerable attention to the connection between color behavior and the brightness state and variability of blazars.
However, earlier statistical analyses were often affected by limited sample sizes and uneven subclass representation~\citep{Ikejiri.11.pasp}. A notable step forward was made by \citet{Negi.22.mn}, who performed the first large-sample analysis based on ZTF DR6 data. 
Building on this, we used a six-year dataset to further test and extend the results obtained from the two-year dataset, while exploring potential trends that may emerge in the segmented time baseline.}

{The results in Section 3.3 indicate a consistent brightness dependence in both the overall and segmented time-window analyses for FSRQs and BL Lacs: brighter states are more likely to exhibit the BWB trend, whereas fainter states are more likely to show the RWB trend.
The fact that the same brightness dependence is found in both BL Lacs and FSRQs suggests that this is a general statistical property of blazar color variability, rather than a subclass-specific peculiarity.}

{The preferential appearance of the BWB trend in brighter states may indicate that flux enhancements are often accompanied by a stronger contribution from the higher-energy part of the synchrotron spectrum. In such cases, the blue band responds more strongly than the red band, naturally leading to a BWB trend.
By contrast, the higher incidence of the RWB trend in fainter states suggests that when the overall flux level is low, the relative contribution of redder emission components becomes more important. For FSRQs, this picture is naturally consistent with a composite scenario in which a relatively blue disk component and a redder variable synchrotron component contribute simultaneously to the observed optical emission. More generally, even without invoking the disk, RWB trend may arise when the lower-frequency synchrotron component dominates the flux variations or when the high-energy blue component fades more rapidly.}

{Our results suggest a further distinction between the two blazar subclasses. Although brighter states in both BL Lacs and FSRQs are more likely to exhibit BWB behavior across different timescales, the dependence on flux variability amplitude appears to be different. In BL Lacs, intervals with larger flux variations are more prone to show BWB trends, which can be naturally understood in a jet-dominated picture: stronger variability likely corresponds to more efficient particle acceleration, enhanced shock activity, or increased Doppler beaming, all of which can strengthen the high-energy synchrotron component and make the optical spectrum bluer as the source brightens. 
By contrast, in FSRQs, stronger flux variability appears more likely to be associated with RWB trends.
This opposite tendency is plausibly related to the additional contribution of the blue thermal accretion disk component in FSRQs. In such a disk+jet system, enhanced variability may often reflect the increasing dominance of a redder non-thermal jet component over a relatively steadier blue disk, thereby favoring RWB trend. However, once the jet becomes sufficiently dominant, FSRQs may also shift toward BWB behavior, as suggested by earlier studies that reported transitions from RWB in low states to BWB in high states or in more jet-dominated phases.}

\section{Conclusions} \label{sec:con}
ZTF provides an unprecedented opportunity to study the long-term variability of AGN through its powerful optical monitoring capabilities. This paper presents a systematic analysis of the brightness and color variability in blazars, based on over six years of quasi-simultaneous observational data for {1149} sources from the ZTF DR22, including {589} BL Lacs and {560} FSRQs. 
We characterize the amplitude of the variability and fractional rms flux variability for each source in the sample, and perform a statistical analysis of both the overall and short-term color behaviors across different subclasses.
Furthermore, we investigate the distribution of brightness {and} variability characteristics across different types of blazars and explore how variability patterns correlate with distinct color behaviors.
\begin{enumerate}
\item 
{Within our six-year optical dataset, BL Lacs on average exhibit greater variability than FSRQs; however, a small subset of FSRQs in our sample shows significantly larger variability amplitudes than the BL Lac population. We emphasize that this should be interpreted as a sample-dependent statistical result rather than a universal trend.}
This can be attributed to the stronger Doppler factors and smaller viewing angles of these FSRQs, which enhance the beaming effect and consequently lead to higher variability.
\item
{Using the six-year dataset, we find that although BL Lacs with negligible host-galaxy contamination generally tend to exhibit toward a BWB trend (14.7\%/68/462) and FSRQs more often show RWB trend (2.3\%/49/560), the low fractions of significant overall trends suggest that long-term blazar color evolution is complex and that short-term spectral variations are largely diluted over extended time baselines.
The source-dependent short-term color behavior also reveals the diversity of physical processes governing blazar evolution.
}
\item 
{Our results further suggest that brighter states in both BL Lacs and FSRQs are more likely to exhibit BWB trend across different timescales, while the dependence on variability amplitude differs between the two subclasses: BL Lacs with a BWB trend show greater variability than those with a RWB trend, whereas FSRQs with a RWB trend exhibit significantly stronger variability than those with a BWB trend, implying that the dominant particle acceleration and emission processes may differ between the two source classes.}
\end{enumerate}

\begin{acknowledgements}

This work was supported by the Development Project of Science and Technology of Jilin Province (grant no. 20260602043RC, 20250102012JC), the Special Project for the Theoretical Basic Research of Changchun Observatory, National Astronomical Observatories, Chinese Academy of Sciences (grant no. Y990000205) and the Fundamental Research Funds for the Central Universities, JLU.
Meng Zhang was supported by the Natural Science Foundation of Shandong Province (grant no. ZR2024QA219).
Ming Zhang is supported by the National Science Foundation of China (12173078).

\end{acknowledgements}

\bibliography{main}{}
\bibliographystyle{aasjournal}

\end{document}